# Superconductivity in $Bi_3O_2S_2Cl$ with Bi–Cl Planar Layers

Bin-Bin Ruan[1, 2], Kang Zhao[1, 2], Qing-Ge Mu[1, 2], Bo-Jin Pan[1, 2], Tong Liu[1, 2], Huai-Xin Yang[1, 2, 3, 4], Jian-Qi Li[1, 2, 3, 4], Gen-Fu Chen[1, 2, 3, 4], and Zhi-An Ren[1, 2, 3, 4, *]

[1] Institute of Physics and Beijing National Laboratory for Condensed Matter Physics, Chinese Academy of Sciences, Beijing 100190, China
[2] School of Physical Sciences, University of Chinese Academy of Sciences, Beijing 100049, China
[3] Collaborative Innovation Center of Quantum Matter, Beijing 100190, China
[4] Songshan Lake Materials Laboratory, Dongguan, Guangdong 523808, China
[*] E-mail: renzhian@iphy.ac.cn

**Abstract**

A quaternary compound $Bi_3O_2S_2Cl$, which consists of novel $[BiS_2Cl]^{2-}$ layers, is reported. It adopts a layered structure of the space group *I*4/*mmm* (No. 139) with lattice parameters: $a$ = 3.927(1) Å, $c$ = 21.720(5) Å. In this compound, bismuth and chlorine atoms form an infinite planar layer, which is unique among the bismuth halides. Superconductivity is observed in both polycrystals and single crystals, and is significantly enhanced in the samples prepared with less sulfur or at higher temperatures. By tuning the content of sulfur, $Bi_3O_2S_2Cl$ can be converted from a semiconductor into a superconductor. The superconducting critical temperature ranges from 2.6 K to 3.5 K. Our discovery of the $[BiS_2Cl]^{2-}$ layer opens another door in searching for the bismuth compounds with novel physical properties.

Bismuth chalcogenides of layered structure have been intensively studied in recent years. For instance, $Bi_2Te_3$ is one of the best performing thermoelectric materials at room temperature.[1-3] More interests were attracted after $Bi_2Se_3$ and $Bi_2Te_3$ were identified as topological insulators.[4,5] Recently, superconductivity was discovered in the $BiCh_2$-based compounds (Ch = S or Se), such as $Bi_4O_4S_3$, $Bi_3O_2S_3$, $REO_{1-x}F_xBiS_2$ (RE = La, Ce, Pr, Nd, Yb), $La_{1-x}M_xOBiS_2$ (M = Ti, Zr, Hf, Th), $Sr_{1-x}La_xFBiS_2$, and $LaO_{1-x}F_xBiSe_2$.[6-15] These compounds are of layered structures where superconductivity was introduced by carrier doping in semiconducting parent compounds, which is quite similar to the cases in the cuprates and iron-based superconductors.[16,17]

All of these superconducting $BiCh_2$-based compounds consist of $[Bi_2Ch_4]^{2-}$ layers separated by other stacking layers, which are basically fluorite-type layers such as $[Bi_2O_2]^{2+}$, $[RE_2O_2]^{2+}$, $[Sr_2F_2]^{2+}$, or the more complicated $[Eu_3F_4]^{2+}$ layers.[6,8,9,18] Large anions, say $[SO_4]^{2-}$ in $Bi_4O_4S_3$ and $S_2^{2-}$ in $Bi_3O_2S_3$, can also be inserted between the layers.[6,7] The tolerance of anion incorporation implies that we might insert other kinds of anions, like halogens into the lattice to form new structures.

The physical properties of the bismuth halides/oxyhalides, on the other hand, were not much studied until recent years when some of them were found to have outstanding photochemical performances. For instance, BiOCl is an efficient photocatalyst to decompose organic pollutants.[19,20] More recently, the bismuth halide perovskites $Cs_3Bi_2X_9$ (X = Cl, Br, or I) were found to have outstanding photoluminescence properties and hence potential applications in solar cells.[21-23] Bismuth halides usually adopt three-dimensional structures as in $BiCl_3$, $Bi_6Cl_7$, and $Bi_7Cl_{10}$,[24-26] while bismuth and halogens form quasi-one-dimensional chains in the cases of BiBr and BiI.[27,28] In many bismuth halides such as $BiI_3$, $Cs_3BiCl_6$, or $Cs_3Bi_2X_9$, the bismuth and halogen atoms form octahedra complexes of $[BiX_6]^{3-}$, which are linked with each other to form the perovskite phases.[21,29,30] However, none of these halides consists of infinite Bi-X planar layers.

BiOCl adopts a layered tetragonal crystal structure in which the $[Bi_2O_2]^{2+}$ layers are separated by the $Cl^-$ anions.[31] Due to the structural similarity with $LaOBiS_2$, one can expect it to be an ideal source to introduce halogens into the $BiCh_2$-based compounds. Here we report the introduction of $Cl^-$ anions into the Bi-O-S system by using BiOCl, which resulted in a layered compound $Bi_3O_2S_2Cl$. Instead of being inserted between the $[Bi_2S_4]^{2-}$ and $[Bi_2O_2]^{2+}$ layers, the Cl atoms form a novel $[BiS_2Cl]^{2-}$ layer. To our knowledge, it is the first compound with this kind of layer. Superconductivity was observed in $Bi_3O_2S_2Cl$, which was introduced by carrier doping caused by the sulfur vacancies.

Figure 1 shows the powder X-ray diffraction (XRD) pattern of polycrystalline $Bi_3O_2S_2Cl$. Except for tiny impurities, all the reflection peaks can be attributed to a new compound $Bi_3O_2S_2Cl$, which crystallizes in the space group of *I*4/*mmm* (No. 139) with the lattice parameters of $a = 3.927(1)$ Å, $c = 21.720(5)$ Å. The most significant impurity peak is around 28.5°, which is attributed to $Bi_2S_3$. The experimental and refinement details can be found in the Supporting Information. The refined structural parameters are summarized in Table 1, and the structure is depicted as the inset of Figure 1. Notice that the occupancy of the sulfur site is about 0.92, indicating sulfur vacancies in the sample. The existence of sulfur vacancies is confirmed by both the energy-dispersive X-ray (EDX) spectrometer (giving an atomic ratio Bi:S:Cl = 3.0:1.6:1.0) (Figure S1) and the inductively coupled plasma (ICP, giving an atomic ratio Bi:S:Cl = 3.00:1.98:1.00) measurements. The loss of sulfur is probably caused by the residue of $Bi_2S_3$, as well as the evaporation of sulfur during the synthesizing procedure.

The transmission electron microscope (TEM) images of different cuts from $Bi_3O_2S_2Cl$ polycrystals are shown in Figure 2 and Figure S2, where the bright spots stand for the positions of large bismuth atoms. The refined positions of the bismuth atoms fit well with the TEM images, both of which suggest a novel $[BiS_2Cl]^{2-}$ layer. This structural feature is distinctly different from the $BiCh_2$-based compounds, such as $LaOBiS_2$, $Bi_4O_4S_3$, $Bi_3O_2S_3$,

Eu$_3$F$_4$Bi$_2$S$_4$, or LaO$_{1-x}$F$_x$BiSe$_2$, all of which consist of [Bi$_2$Ch$_4$]$^{2-}$ layers.[6-15,18] Notice that for the [Bi$_2$S$_4$]$^{2-}$ layers in LaOBiS$_2$, the in-plane sulfur has the priority to be substituted by selenium.[33] Therefore, it is likely that Cl takes the intra-layer sites while S takes the inter-layer ones of the [BiS$_2$Cl]$^{2-}$ layer. This assumption is further backed by the impossibility of mutual doping between S and Cl in our experiments. The geometry optimization and bond valence sum calculation support this assumption too. (see Supporting Information) The unforeseen [BiS$_2$Cl]$^{2-}$ layer, which is an infinite square planar layer of Bi-Cl decorated with S anions, is unique among the bismuth compounds.

Figure 3(a) shows the temperature dependence of resistivity for the polycrystalline Bi$_3$O$_2$S$_2$Cl sample under zero magnetic field. The resistivity shows a metallic behavior below 300 K, where a shallow minimum at around 15 K is observed. This pit-like feature of resistivity was also observed in other bismuth sulfides, such as Bi$_2$S$_3$ and K$_2$Bi$_8$S$_{13}$.[34,35] It results from the freezing of carriers at lower temperatures. Besides, the absolute value of resistivity at 300 K is significantly larger than typical metals, while comparable with those in doped bismuth chalcogenide semiconductors such as CsBi$_4$Te$_6$, LaO$_{1-x}$F$_x$BiS$_2$, or Sr$_{1-x}$La$_x$FBiS$_2$.[8,9,36] A drop of resistivity is observed below the superconducting critical temperature $T_c$ (2.6 K), indicating the occurrence of superconductivity. The details of the superconducting transition are demonstrated in Figure 3(b). The relatively broad transitions should be attributed to the inhomogeneity of different doping levels in the sample. Various magnetic fields were applied to investigate the superconducting properties. As the magnetic field increases, $T_c$ decreases and the superconducting transition broadens, indicating a suppression of superconductivity by magnetic field. The superconductivity is fully suppressed by a field of 1 T above 1.8 K.

We measured the Hall effects of polycrystalline Bi$_3$O$_2$S$_2$Cl to clarify the carrier's type as well as its concentration in the sample. The measurement was carried out on a thin rectangle cut from the polycrystalline sample. The dimension of the sample was about $8 \times 4 \times 1$ mm$^3$. The field dependence of Hall resistivity ($\rho_{xy}$) was measured at various temperatures from 10 K to 275 K. At each temperature, $\rho_{xy}$ was found to be linear with fields up to 4 T, the highest field applied. The field dependences of $\rho_{xy}$ at various temperatures are taken as examples and shown in Figure 3(c). The linearity of the Hall effect, which is again distinctly different from that of LaOBiS$_2$, Bi$_4$O$_4$S$_3$, and Bi$_3$O$_2$S$_3$, suggests a different band structure near the Fermi surface. (see Supporting Information)[7,37-39] This difference may arise from the dissimilarity between [BiS$_2$Cl]$^{2-}$ and [Bi$_2$S$_4$]$^{2-}$ layers. Since the Hall coefficient $R_H$ is field independent, it is easily determined by the slope of $\rho_{xy}$ versus $\mu_0 H$. The values of $R_H$ are negative at all temperatures, suggesting that electrons serve as the dominant carriers. The charge carrier concentration is calculated by $n = 1/(eR_H)$, giving $9.34 \times 10^{19}$ cm$^{-3}$ at 275 K. This value is comparable with that of Bi$_4$O$_4$S$_3$ (~ $7.28 \times 10^{19}$ cm$^{-3}$ at 300 K) yet is much lower than that of typical metals (~ $10^{22}$ cm$^{-3}$).[40] The carrier density corresponds to 0.016 electrons per formula at 275 K, which could be easily introduced by a tiny amount of sulfur vacancies. As shown in Figure 3(d), the carrier concentration varies little in the whole temperature range while decreases a little at low temperatures, indicating Bi$_3$O$_2$S$_2$Cl to be a heavily doped semiconductor.

To verify the semiconducting-superconducting transition in Bi$_3$O$_2$S$_2$Cl, intentional carrier doping was carried out. We conducted many kinds of doping such as the fluorine substitution of oxygen, the selenium substitution of sulfur, or the doping of element vacancies. Unfortunately, only the doping of sulfur vacancies was successful, which led to changes of the lattice parameters and transport properties. With an increasing of sulfur vacancies, the X-ray diffraction peaks of Bi$_3$O$_2$S$_x$Cl shift to higher degrees, indicating a shrinkage of the lattice. (Figure S3)

The resistivity of Bi$_3$O$_2$S$_x$Cl ($x$ = 2.1, 2.0, 1.95, 1.9) samples with different sulfur contents is shown in Figure 4(a), where $x$ stands for the nominal value of the sulfur quantity. Here we successfully obtained the non-superconducting sample of the nominal composition of $Bi_3O_2S_{2.1}Cl$ which indeed demonstrated a semiconducting behavior below 100 K. This sample could be considered as the "parent compound" of the $Bi_3O_2S_2Cl$ superconductors. The conductivity of the sample increases with the introduction of sulfur vacancies, and superconductivity of $T_c$ = 3.5 K was also observed. As revealed in Figure 4(b), the diamagnetic signals grow larger above 1.8 K. In other words, doping of sulfur vacancies enhances the superconducting phase in the sample, which could also be the origin of superconductivity in the pristine (not intentionally doped) $Bi_3O_2S_2Cl$.

In some batches of the $Bi_3O_2S_2Cl$ polycrystal growth, we were able to collect a few pieces of plate-like single crystals on the sample surfaces, including brownish and transparent ones. XRD measurement reveals the brownish ones to be $Bi_3O_2S_2Cl$, while the transparent ones are BiOCl. The typical size of $Bi_3O_2S_2Cl$ single crystals is about 0.08×0.06×0.01 $mm^3$, as shown in the inset of Figure 5(a). X-ray diffraction of the single crystals only consists of (00$l$) reflections ($l$ = 2, 4, 6, 8…), indicating the surface of the single crystals to be perpendicular to the $c$-axis. The reflections are also consistent with the body-centered space group of $Bi_3O_2S_2Cl$. According to the (00$l$) peaks, the lattice parameter $c$ is calculated to be 21.70 Å, which is slightly smaller than that of the polycrystals. The magnetic susceptibility of the single crystals was measured with magnetic field parallel to the plane of the sample. Superconductivity with a $T_c$ = 2.8 K is also observed in the single-crystalline $Bi_3O_2S_2Cl$ sample. The diamagnetic signals are both significant in the ZFC and the FC runs, verifying the superconductivity. Notice that the diamagnetic signal is much larger than those of the polycrystalline samples above 1.8 K.

The relatively higher growing temperatures (compared to the polycrystalline ones, see Supporting Information) tended to introduce more sulfur vacancies, which resulted in a smaller $c$-axis and consequentially a larger shielding fraction of superconductivity in the single crystals. Carrier doping was realized either by the existence of site vacancies, or by the formation of stacking faults (such as the BiOCl-like stacking faults) caused by the lack of sulfur. Further studies should be carried out to obtain more and larger single crystals to elucidate the electronic consequences of sulfur vacancies therein.

In summary, a new layered compound $Bi_3O_2S_2Cl$ was synthesized via solid-state reactions. As revealed by the Rietveld refinement and TEM images, $Bi_3O_2S_2Cl$ adopts a layered-type structure built with alternatively stacking of $[Bi_2O_2]^{2+}$ and $[BiS_2Cl]^{2-}$ layers. It marks the first compound with the infinite Bi-Cl square-planar layers. Electrical transport measurements reveal $Bi_3O_2S_2Cl$ to be an n-type semiconductor, in which superconductivity can be introduced by doping of the sulfur vacancies. The $[BiS_2Cl]^{2-}$ layer is a novel building block in the layered compounds, which provides another option for the search of bismuth materials.

This work was supported by the National Natural Science Foundation of China (Grant Nos. 11474339, and 11774402), the National Basic Research Program of China (Grant No. 2016YFA0300301), and the Youth Innovation Promotion Association of the Chinese Academy of Sciences.

**Figure 1.** Powder X-ray diffraction pattern of $Bi_3O_2S_2Cl$ with its Rietveld refinement. The inset shows the crystal structure obtained from the refinement, which is drawn with VESTA.[32]

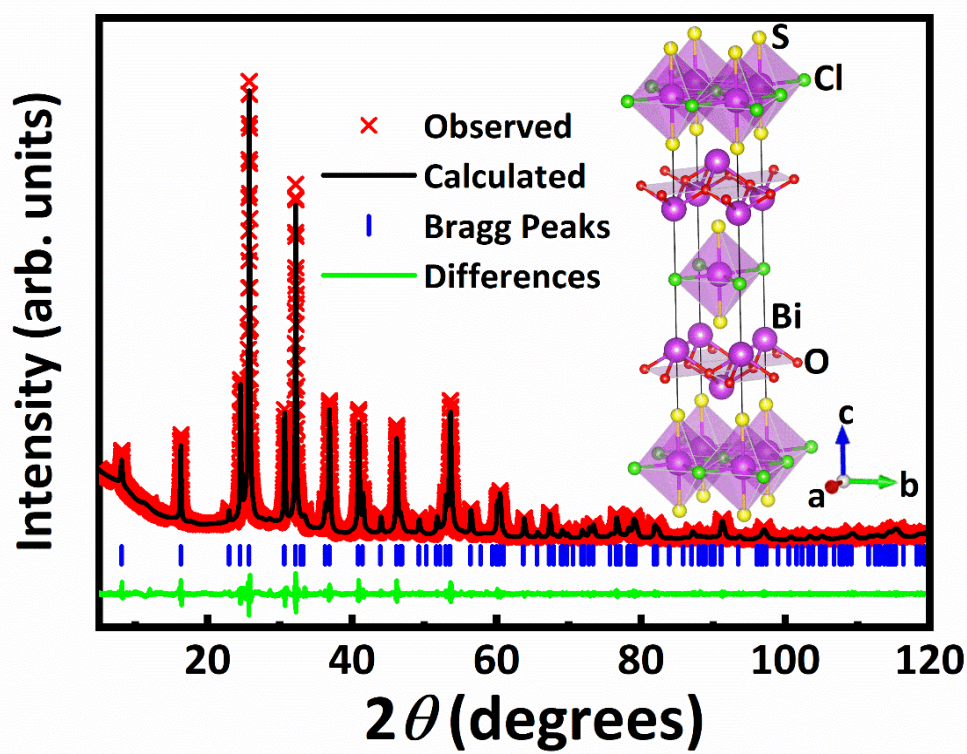

**Figure 2.** (a) Appearance of the polycrystalline $Bi_3O_2S_2Cl$ under scanning electron microscope (SEM). (b), (c), (d) TEM images observed from the directions of [001], [100], and [110], respectively. Insets are the sketches of the refined crystal structure for comparison.

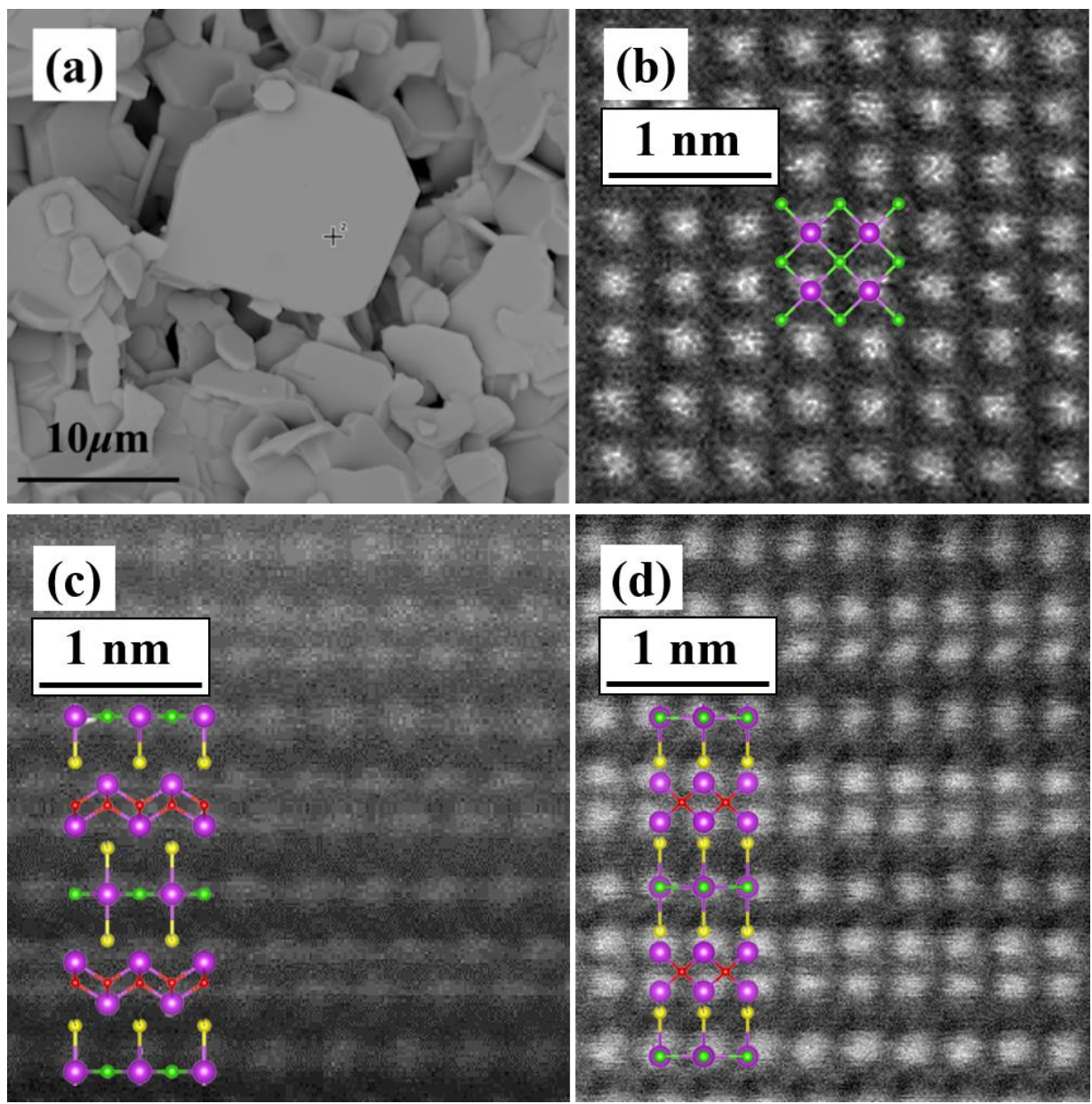

**Figure 3.** (a) Temperature dependence of resistivity of polycrystalline $Bi_3O_2S_2Cl$. (b) Resistivity under various magnetic fields. (c) Hall resistivity under various temperatures. (d) Carrier concentration calculated from Hall measurement.

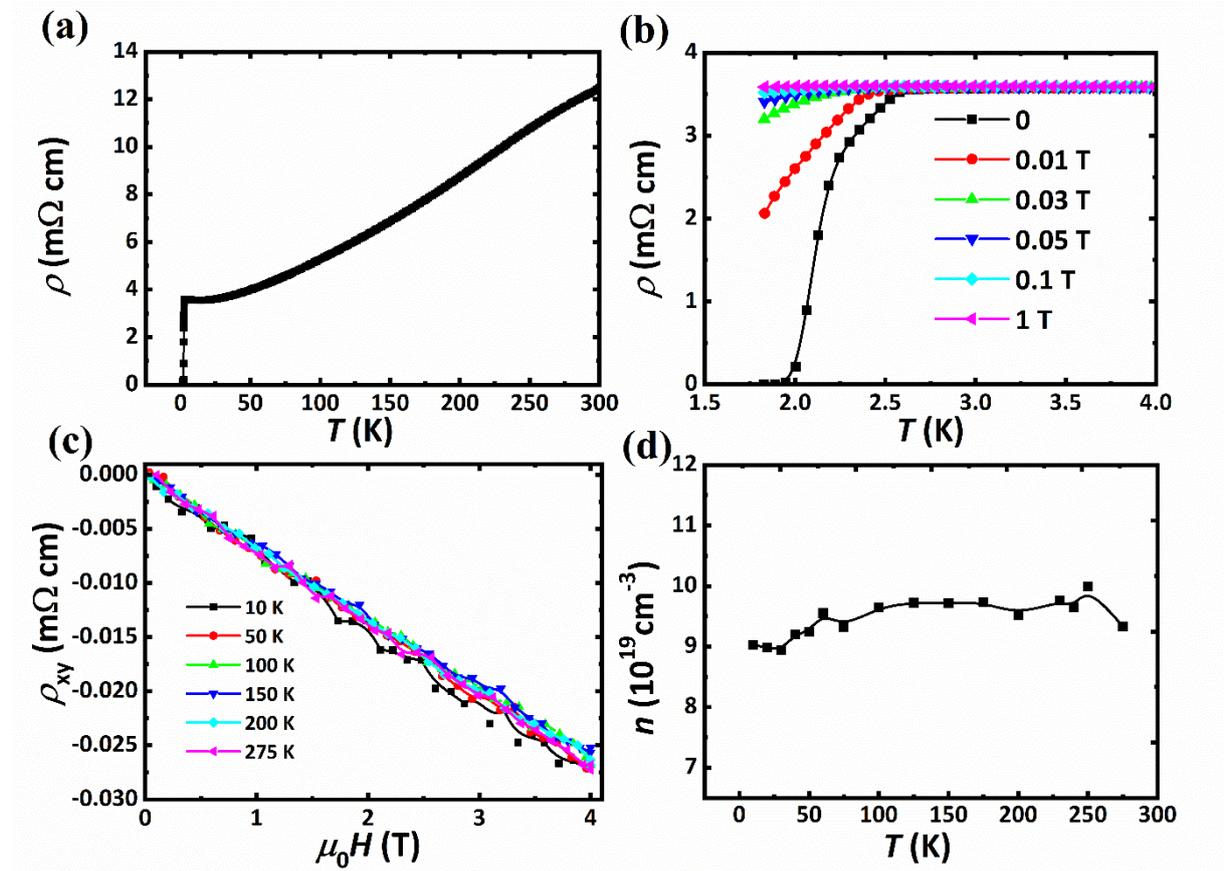

**Figure 4.** (a) Temperature dependence of resistivity and (b) magnetic susceptibility of the doped samples of nominal compositions of $Bi_3O_2S_xCl$ ($x$ = 2.1, 2.0, 1.95, 1.9).

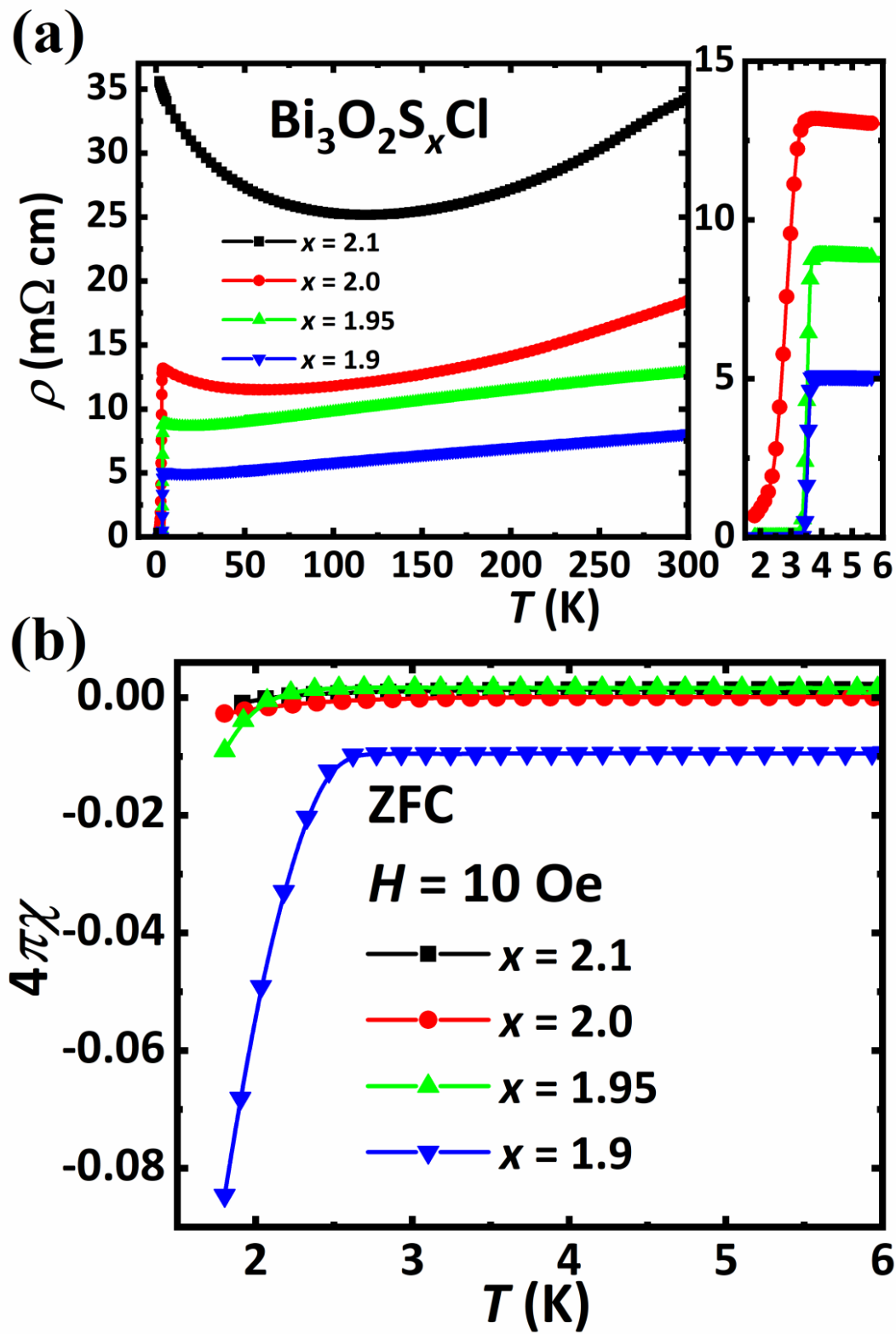

**Figure 5.** (a) Magnetic susceptibility of $Bi_3O_2S_2Cl$ single crystals. The inset shows the appearance of the single crystals under an optical microscope. (b) X-ray diffraction pattern of the surface of the single crystals. The labeled peaks belong to the $Bi_3O_2S_2Cl$ phase, while small amount of BiOCl is marked with the asterisk.

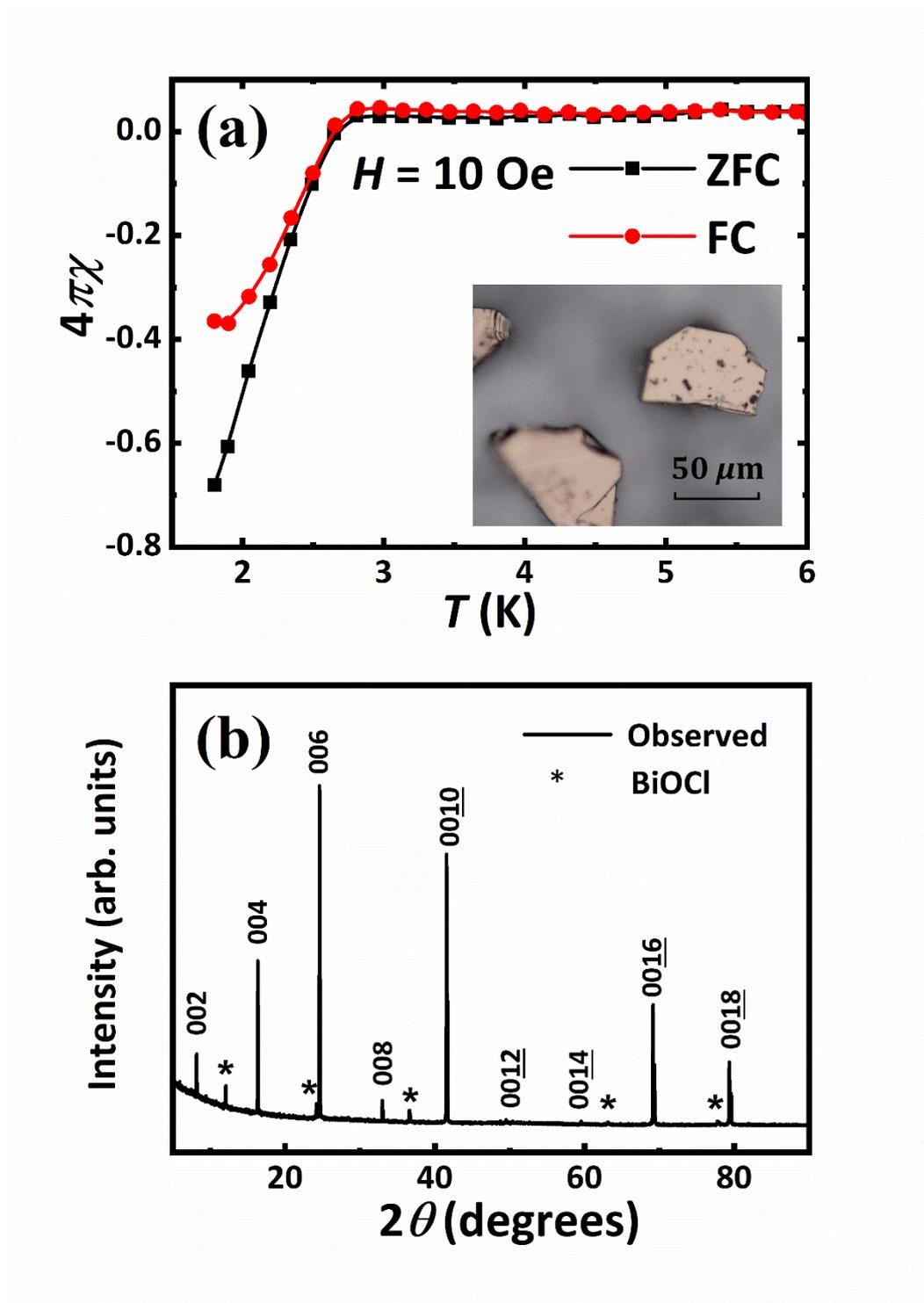

**Table 1.** structural parameters obtained by Rietveld refinement.

| atom | site | x | y | z | occupancy | $U_{iso}$ (Å$^2$) |
| --- | --- | --- | --- | --- | --- | --- |
| Bi1 | 2a | 0 | 0 | 0 | 1.00(3) | 0.056(1) |
| Bi2 | 4e | 0 | 0 | 0.3108(1) | 0.99(2) | 0.024(1) |
| S | 4e | 0 | 0 | 0.1298(5) | 0.92(3) | 0.016(2) |
| O | 4d | 0.5 | 0 | 0.25 | 1.00(1) | 0.012(2) |
| Cl | 2b | 0.5 | 0.5 | 0 | 1.00(1) | 0.048(3) |